\documentclass[a4paper,10pt,twoside]{cpc-hepnp}

\usepackage{multicol}
\usepackage{graphicx}
\usepackage{amssymb,bm,mathrsfs,bbm,amscd}
\usepackage[tbtags]{amsmath}
\usepackage{lastpage}

\begin{document}

\title{A Test of CPT Symmetry in $K^0$ vs $\bar{K}^0
\rightarrow \pi^+\pi^-\pi^0$ Decays\thanks{Supported in part by
the National Natural Science Foundation of China (10425522 and
10875131) and in part by the Ministry of Science and Technology of
China (2009CB825207).} }

\author{%
XING Zhi-Zhong$^{1,2;1)}$\email{xingzz@ihep.ac.cn}}

\maketitle

\address{
1~(Institute of High Energy Physics, Chinese Academy of Sciences,
Beijing 100049, China)\\
2~(Theoretical Physics Center for Science Facilities,
Chinese Academy of Sciences, Beijing 100049, China)\\
}

\begin{abstract}
I show that the CP-violating asymmetry in $K^0$ vs $\bar K^0
\rightarrow \pi^+\pi^-\pi^0$ decays differs from that in $K^{}_{\rm
L} \rightarrow \pi^+\pi^-$, $K^{}_{\rm L} \rightarrow \pi^0\pi^0$ or
the semileptonic $K^{}_{\rm L}$ transitions, if there exists CPT
violation in $K^0$-$\bar{K}^0$ mixing. A delicate measurement of
this difference at a super flavor factory (e.g., the $\phi$ factory)
will provide us with a robust test of CPT symmetry in the neutral
kaon system.
\end{abstract}



\footnotetext[0]{\hspace*{-2em}\small\centerline{\thepage\ --- \pageref{LastPage}}}%

\begin{multicols}{2}

\section{The motivation}

The CPT theorem claims that a Lorentz-invariant local quantum field
theory with a Hermitian Hamiltonian must have CPT symmetry
\cite{CPT}. It is so far so good, because there is no convincing
experimental hint at CPT violation \cite{PDG08}. The breaking of CPT
symmetry, as expected in some ``exotic" scenarios of new physics
beyond the standard model (e.g., string theory) \cite{Alan}, would
be a big deal. In any case, much more experimental tests of this
theorem are desirable.

The $K^0$-$\bar{K}^0$ mixing system has been playing an important
role in particle physics for testing fundamental symmetries (such as
CP, T and CPT) and examining conservation laws (such as $\rm\Delta S
= \Delta Q$). The existing experimental evidence for CPT invariance
in the mixing and decays of neutral kaon mesons remains rather poor
\cite{PDG08}: it is not excluded that the strength of CPT-violating
interactions could be as large as about ten percentage of that of
CP-violating interactions. This unsatisfactory situation will be
improved in the near future, in particular after a variety of more
delicate measurements are carried out at a super flavor factory
\cite{Hitlin} (e.g., the $\phi$ factory \cite{Phi}).

There are several possibilities of probing CPT violation in
$K^0$-$\bar{K}^0$ mixing with the decays of $K^{}_{\rm S}$ and
$K^{}_{\rm L}$ mesons into the two-pion and (or) the semileptonic
states \cite{PDG08}. A different approach towards testing CPT
symmetry, with the help of neutral kaon decays into the three-pion
states, has also been pointed out in Ref. \cite{Xing00}. The idea is
simply that the CP-violating effect induced by $K^0$-$\bar K^0$
mixing in $K^0$ vs $\bar K^0 \rightarrow \pi^+\pi^-\pi^0$
transitions should not be identical to that in $K^{}_{\rm
L}\rightarrow \pi^+\pi^-$, $K^{}_{\rm L} \rightarrow \pi^0\pi^0$ or
the semileptonic $K^{}_{\rm L}$ decays, if CPT symmetry is broken.
Thus a careful comparison between these two types of CP-violating
effects may provide us with a robust test of CPT invariance in
$K^0$-$\bar K^0$ mixing.

An unfortunate fact is that no attention has so far been paid to the
method advocated in Ref. \cite{Xing00}. In this talk, which is more
or less an advertisement, I shall explain why a test of CPT symmetry
is possible by measuring the time-dependent CP-violating asymmetry
between $K^0(t) \rightarrow \pi^+\pi^-\pi^0$ and $\bar{K}^0(t)
\rightarrow \pi^+\pi^-\pi^0$ decays. My result is hopefully useful
for the upcoming experiments of kaon physics.

\section{The idea}

Let me outline the main idea. The mass eigenstates of $K^0$ and
$\bar K^0$ can in general be written as
\begin{eqnarray}
|K^{}_{\rm S} \rangle & = & \frac{1}{\sqrt{|p^{~}_1|^2 +
|q^{~}_1|^2}} \left ( p^{~}_1 |K^0\rangle + q^{~}_1 |\bar{K}^0
\rangle \right ) \; , \nonumber \\
|K^{}_{\rm L} \rangle & = & \frac{1}{\sqrt{|p^{~}_2|^2 +
|q^{~}_2|^2}} \left ( p^{~}_2 |K^0\rangle - q^{~}_2
|\bar{K}^0\rangle \right ) \; ,
\end{eqnarray}
in which $p^{~}_i$ and $q^{~}_i$ (for $i=1,2$) are complex mixing
parameters. Note that $p^{~}_1 =p^{~}_2$ and $q^{~}_1 = q^{~}_2$
follow from CPT invariance \cite{Lee}. The traditional
characteristic quantities of CP violation in the $K^0$-$\bar{K}^0$
mixing system \cite{PDG08}, $\eta_{+-}$, $\eta_{00}$ and
$\delta^{}_{\rm L}$, are all related to $K^{}_{\rm L}$ decays and
thus the $(p^{~}_2, q^{~}_2)$ parameters. For example,
\begin{equation}
\delta^{}_{\rm L} \;\; \equiv \;\; \frac{|p^{~}_2|^2 -
|q^{~}_2|^2}{|p^{~}_2|^2 + |q^{~}_2|^2} \;
\end{equation}
in the absence of $\rm\Delta S = - \Delta Q$ interactions. A
measurement of CP violation associated with
\begin{equation}
\delta^{}_{\rm S} \;\; = \;\; \frac{|p^{~}_1|^2 -
|q^{~}_1|^2}{|p^{~}_1|^2 + |q^{~}_1|^2} \;
\end{equation}
has been assumed to be extremely difficult, if not impossible, due
to the rapid decay of the $K^{}_{\rm S}$ meson to the two-pion
state or the semileptonic state. Nevertheless, I shall show that
$\delta^{}_{\rm S}$ can be measured from the rate asymmetry of
$K^0$ and $\bar{K}^0$ mesons decaying into the three-pion state
$\pi^+\pi^-\pi^0$. The difference between $\delta^{}_{\rm S}$ and
$\delta^{}_{\rm L}$ signifies CPT violation in $K^0$-$\bar{K}^0$
mixing. This point can be seen more clearly if one adopts the
popular $(\epsilon, \delta)$ parameters to describe CP- and
CPT-violating effects in the $K^0$-$\bar K^0$ mixing system
\cite{PDG08}:
\begin{eqnarray}
p^{~}_1 & = & 1 + \epsilon + \delta \; , \nonumber \\
p^{~}_2 & = & 1 + \epsilon - \delta \; , \nonumber \\
q^{~}_1 & = & 1 - \epsilon - \delta \; , \nonumber \\
q^{~}_2 & = & 1 - \epsilon + \delta \; .
\end{eqnarray}
Then
\begin{eqnarray}
\delta^{}_{\rm L} & = & 2 \left ( {\rm Re} \epsilon ~ - ~ {\rm Re}
\delta \right
) \; , \nonumber \\
\delta^{}_{\rm S} & = & 2 \left ( {\rm Re} \epsilon ~ + ~ {\rm Re}
\delta \right ) \; .
\end{eqnarray}
It turns out that $\delta^{}_{\rm S} - \delta^{}_{\rm L} = 4 {\rm
Re}\delta$ is a clear signature of CPT violation \cite{Xing00}.

Let me quote two typical experimental constraints on the
CPT-violating parameter $\delta$ in $K^0$-$\bar{K}^0$ mixing: ${\rm
Re}\delta = (2.9 \pm 2.6^{}_{\rm stat} \pm 0.6^{}_{\rm syst}) \times
10^{-4}$ obtained by the CPLEAR Collaboration \cite{CPL98} and ${\rm
Im}\delta = (0.4 \pm 2.1) \times 10^{-5}$ obtained by the KLOE
Collaboration \cite{KLOE}. A systematic analysis of the CP- and
CPT-violating parameter space has already been done by the Particle
Data Group in Ref. \cite{PDG08}.

\section{The approach}

The CP eigenvalue for the $\pi^+\pi^-\pi^0$ final state is given
by $(-1)^{l+1}$, where $l$ is the relative angular momentum
between $\pi^+$ and $\pi^-$. Since the sum of the masses of three
pions is close to the kaon mass, the pions have quite low kinetic
energy $E^{}_{\rm CM}(\pi)$ in the kaon rest-frame, and the states
with $l>0$ are suppressed by the centrifugal barrier
\cite{CPLEAR}. Thus the $K^{}_{\rm L}$ meson decays dominantly
into the kinematics-favored ($l=0$) and CP-allowed (CP = $-$1)
$\pi^+\pi^-\pi^0$ state. The decay amplitude of $K^{}_{\rm
S}\rightarrow \pi^+\pi^-\pi^0$ consists of both the
kinematics-suppressed ($l=1$) but CP-allowed (CP=+1) component,
and the kinematics-favored ($l=0$) but CP-forbidden (CP=$-$1)
component. This implies an interesting Dalitz-plot distribution
for the $K^{}_{\rm S}\rightarrow \pi^+\pi^-\pi^0$ transition: it
is symmetric with respect to $\pi^+$ and $\pi^-$ for the
CP-violating amplitude, but anti-symmetric for the CP-conserving
amplitude. Let the ratio of $K^{}_{\rm S}$ and $K^{}_{\rm L}$
decay amplitudes be
\begin{equation}
\eta^{~}_{+-0} \;\; = \;\; \frac{A (K^{}_{\rm S} \rightarrow
\pi^+\pi^-\pi^0)} {A(K^{}_{\rm L} \rightarrow \pi^+\pi^-\pi^0)} \; .
\end{equation}
It is clear that $\eta^{~}_{+-0}$ depends only upon the CP-violating
component of $A( K^{}_{\rm S} \rightarrow \pi^+\pi^-\pi^0)$, if data
are integrated over the whole Dalitz plot \cite{CPLEAR,Nakada98}.
The time-dependent rates for the initially pure $K^0$ and $\bar K^0$
states decaying into $\pi^+\pi^-\pi^0$, denoted by ${\cal R}(t)$ and
$\bar{\cal R}(t)$ respectively, can be calculated with the help of
Eqs. (1) and (6). I arrive at \cite{Xing00}
\end{multicols}
\ruleup
\begin{eqnarray}
{\cal R}(t) & \propto & \left[ |q^{~}_1|^2 + |q^{~}_2|^2
|\eta_{+-0}|^2 e^{-\Delta \Gamma t} + 2 {\rm Re} \left(q^*_1 q^{~}_2
\eta_{+-0} e^{i \Delta m t} \right) e^{-\Delta \Gamma t/2} \right]
\; ,
\nonumber \\
\bar{\cal R}(t) & \propto & \left[ |p^{~}_1|^2 + |p^{~}_2|^2
|\eta_{+-0}|^2 e^{-\Delta \Gamma t} - 2 {\rm Re} \left(p^*_1 p^{~}_2
\eta_{+-0} e^{i \Delta m t} \right) e^{-\Delta \Gamma t/2} \right]
\; ,
\end{eqnarray}
\ruledown \vspace{0.7cm}
\begin{multicols}{2}
\hspace{-0.65cm} where $\Delta m >0$ and $\Delta \Gamma >0$ denote
the mass difference and the width difference of $K^{}_{\rm S}$ and
$K^{}_{\rm L}$ mesons, respectively. To a good degree of accuracy, I
obtain the following CP-violating asymmetry:
\end{multicols}
\ruleup
\begin{eqnarray}
{\cal A}(t) \;\; \equiv \;\; \frac{\bar{\cal R}(t) - {\cal R}(t)}
{\bar{\cal R}(t) + {\cal R}(t)} & = & \delta^{}_{\rm S} \; - \; 2
e^{-\Delta \Gamma t/2} \left[ {\rm Re}\eta^{~}_{+-0} \cos (\Delta m
t) - {\rm
Im}\eta^{~}_{+-0} \sin (\Delta m t) \right] \xi \nonumber \\
&& \;\;\;\;\;\;\; - \; 2 e^{-\Delta \Gamma t/2} \left[ {\rm
Re}\eta^{~}_{+-0} \sin (\Delta m t) + {\rm Im}\eta^{~}_{+-0} \cos
(\Delta m t) \right] \zeta \; ,
\end{eqnarray}
\ruledown \vspace{0.7cm}
\begin{multicols}{2}
\hspace{-0.65cm} in which
\end{multicols}
\ruleup
\begin{eqnarray}
\xi & = & \frac{{\rm Re} ( p^{~}_1 p^*_2 + q^{~}_1 q^*_2
)}{|p^{~}_1|^2 + |q^{~}_1|^2} \; = \; 1 + {\cal O}(|\epsilon|^2) +
{\cal O}(|\delta|^2) + {\cal O}({\rm Re}(\epsilon \delta^*)) \; , \nonumber \\
\zeta & = & \frac{{\rm Im} ( p^{~}_1 p^*_2 + q^{~}_1 q^*_2
)}{|p^{~}_1|^2 + |q^{~}_1|^2} \; = \; {\cal O} ({\rm Im}(\epsilon
\delta^*)) \; .
\end{eqnarray}
\ruledown \vspace{0.7cm}
\begin{multicols}{2}
\hspace{-0.65cm} It is obvious that $\delta^{}_{\rm S}$ can be
determined through the measurement of ${\cal A}(t)$. In particular,
the relationship $\displaystyle\lim_{t\rightarrow \infty} {\cal
A}(t) = \delta^{}_{\rm S}$ holds.

As I have emphasized, the difference between $\delta^{}_{\rm S}$ and
$\delta^{}_{\rm L}$ hints at CPT violation in $K^0$-$\bar{K}^0$
mixing. If $|{\rm Re}\delta| / {\rm Re}\epsilon \sim 0.1$, then the
difference $\delta^{}_{\rm S} - \delta^{}_{\rm L} = 4 {\rm
Re}\delta$ can be as large as $0.4 {\rm Re} \epsilon \sim 6.6 \times
10^{-4}$ in magnitude, where the experimental value ${\rm
Re}\epsilon \approx 1.65 \times 10^{-3}$ has been used \cite{PDG08}.
Since both $\epsilon$ and $\delta$ are small quantities, it turns
out that $\xi \approx 1$ and $\zeta \approx 0$ are good
approximations. Eq. (8) is therefore simplified to
\end{multicols}
\ruleup
\begin{equation}
{\cal A}(t) \;\; = \;\; \delta^{}_{\rm S} \; - \; 2 e^{-\Delta
\Gamma t/2} \left [ {\rm Re}\eta^{~}_{+-0} \cos (\Delta m t) ~ - ~
{\rm Im}\eta^{~}_{+-0} \sin (\Delta m t) \right ] \; .
\end{equation}
\ruledown \vspace{0.7cm}
\begin{multicols}{2}
\hspace{-0.65cm} In the neglect of CPT violation, namely,
$\delta^{}_{\rm S} = 2 {\rm Re}\epsilon$, Eq. (10) can simply
reproduce the result obtained in Ref. \cite{CPLEAR}. For
illustration, I plot the behavior of ${\cal A}(t)$ in Fig. 1, in
which $\delta^{}_{\rm S} = 3\times 10^{-3}$ and $|\eta_{+-0}| =
5\times 10^{-3}$ have typically been input. One may observe that
${\cal A}(t)$ approaches $\delta^{}_{\rm S}$ for $t \geq 5
\tau^{~}_{\rm S}$ and reaches $\delta^{}_{\rm S}$ if $t \geq 10
\tau^{~}_{\rm S}$, where $\tau^{~}_{\rm S}$ is the mean lifetime of
the $K^{}_{\rm S}$ meson. This implies a certain feasibility to
determine $\delta^{}_{\rm S}$ from the time-dependent CP-violating
asymmetry ${\cal A}(t)$.
\begin{center}
\includegraphics[bb = 210 540 440 770,scale=0.7]{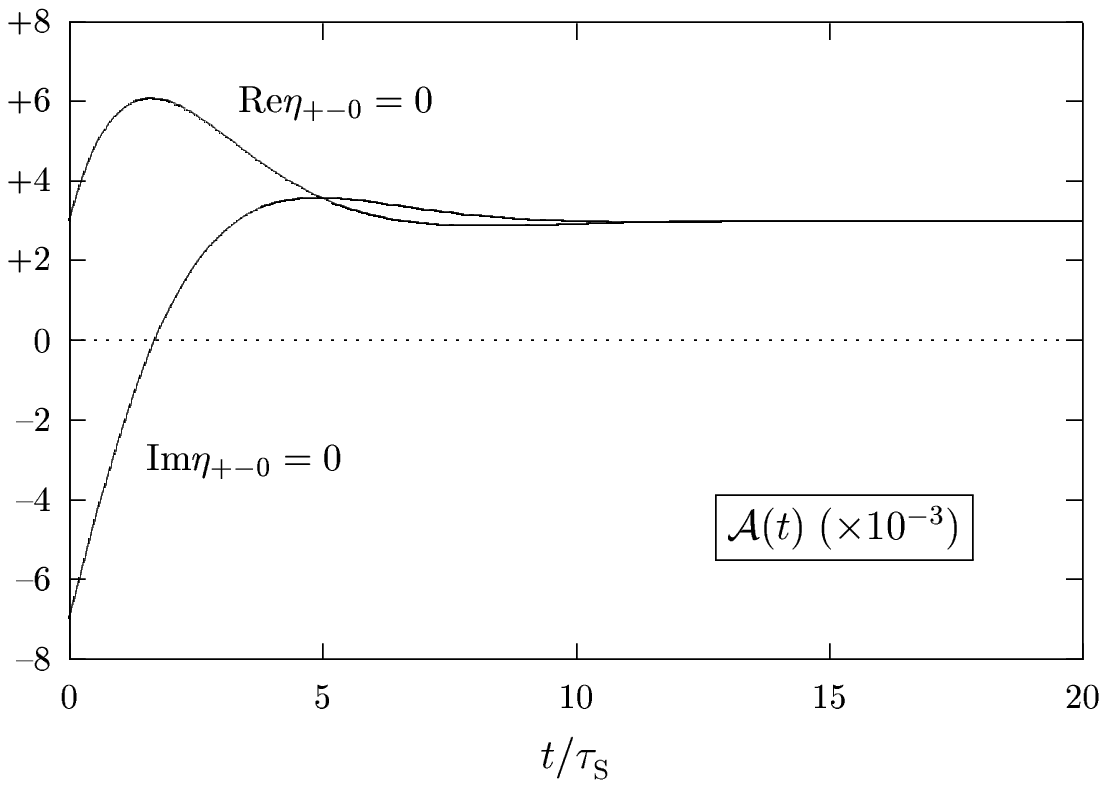}
\figcaption{An illustrative plot for the CP-violating asymmetry
${\cal A}(t)$ with the typical inputs $\delta^{}_{\rm S} = 3 \times
10^{-3}$ and $|\eta^{}_{+-0}| = 5 \times 10^{-3}$ \cite{Xing00}.}
\end{center}

\section{The discussion}

In the above analysis I have taken an integration over the whole
Dalitz plot, such that $\eta^{}_{+-0}$ solely contains the
CP-violating part of $A(K^{}_{\rm S}\rightarrow \pi^+\pi^-\pi^0)$.
To look at the CP-conserving component of $A(K^{}_{\rm S}\rightarrow
\pi^+\pi^-\pi^0)$, one may study the phase-space regions $E^{}_{\rm
CM}(\pi^+) > E^{}_{\rm CM}(\pi^-)$ and $E^{}_{\rm CM}(\pi^+) <
E^{}_{\rm CM}(\pi^-)$ separately \cite{CPLEAR}. In this case the
corresponding CP-violating asymmetries between $\bar{\cal R}(t)$ and
${\cal R}(t)$ take the same form as ${\cal A}(t)$ in Eq. (8) or Eq.
(10), but $\eta^{}_{+-0}$ should be replaced by $(\eta^{}_{+-0} \pm
\lambda)$, where $\lambda$ denotes the CP-conserving contribution to
the ratio of $K^{}_{\rm S}$ and $K^{}_{\rm L}$ decay amplitudes
\cite{CPLEAR}. Certainly, the CP-violating parameter $\delta^{}_{\rm
S}$ can still be determined from measuring the time dependence of
the relevant decay rate asymmetries.

An accurate measurement of $\delta^{}_{\rm S}$ from $K^0$ vs $\bar
K^0 \rightarrow \pi^+\pi^-\pi^0$ should be feasible at the $\phi$
factory, where a huge amount of $K^0\bar K^0$ events can be
coherently produced \cite{Phi}. Choosing the semileptonic decay of
one kaon to tag the flavor of the other kaon decaying into
$\pi^+\pi^-\pi^0$ on the $\phi$ resonance, one should be able to
construct the time-dependent rate asymmetry between $K^0(t)
\rightarrow \pi^+\pi^-\pi^0$ and $\bar K^0(t) \rightarrow
\pi^+\pi^-\pi^0$ decays in a way similar to Eq. (8). It is also
expected that other super flavor factories may measure
$\delta^{}_{\rm S}$ and $\delta^{}_{\rm L}$ to a good degree of
accuracy.

Note that Lorentz invariance has been taken for granted in what I
have discussed. As pointed out by Greenberg \cite{Greenberg}, ``If
CPT invariance is violated in an interacting quantum field theory,
then that theory also violates Lorentz invariance". In my
discussions, the dependence of the CPT-violating parameter $\delta$
on the sidereal time should in general be considered, since CPT
violation may simultaneously imply a violation of Lorentz symmetry
in the neutral kaon system. For simplicity, here I take $\delta$ to
be a constant by assuming that the boost parameters of both $K^0$
and $\bar{K}^0$ are small and the corresponding Lorentz-violating
effect is rotationally invariant in the laboratory frame
\cite{Kos98}. In this approximation, my results are essentially
valid as the averages over the sidereal time, such that the effect
of Lorentz violation due to the direction of motion is negligible.

Finally, I like to mention that different approaches have been
discussed to test CPT symmetry in $D^0$-$\bar{D}^0$,
$B^0_d$-$\bar{B}^0_d$ or $B^0_s$-$\bar{B}^0_s$ mixing \cite{Xing94}.
The idea presented here cannot directly be applied to those heavy
neutral-meson systems. In this sense, it represents a unique way
applicable in the $K^0$-$\bar{K}^0$ mixing system to test the CPT
theorem.

\section{The conclusion}

To conclude, the CP-violating effect induced by $K^0$-$\bar K^0$
mixing in $K^0$ vs $\bar K^0 \rightarrow \pi^+\pi^-\pi^0$ decays is
possible to deviate to some extent from that in $K^{}_{\rm L}
\rightarrow \pi\pi$ or the semileptonic $K^{}_{\rm L}$ transitions
due to the violation of CPT symmetry. Measuring or constraining this
tiny difference may serve as a robust test of CPT invariance in the
neutral kaon system.

I would like to thank Changzheng Yuan and the organizing committee
of PHIPSI09 for giving me this opportunity to present an ``old''
idea for the future.

\end{multicols}

\vspace{-2mm} \centerline{\rule{80mm}{0.1pt}} \vspace{2mm}

\begin{multicols}{2}

\end{multicols}

\clearpage

\end{document}